\documentclass[11pt,a4paper]{article}
\pdfoutput=1
\usepackage{bbold,bm}
\usepackage{jinstpub}
\usepackage{lineno}
\usepackage{multirow}
\usepackage{rotating}
\usepackage{notes2bib}
\usepackage{longtable} 
\usepackage{epstopdf}
\usepackage{xcolor}
\usepackage{lipsum}
\usepackage{url}
\usepackage{verbatimbox}

\newcommand{\MJD}{\textsc {Majorana Demonstrator}}

\newcommand{\Dem}{\textsc{Demonstrator}}
\newcommand{\nonubb}{$0\nu\beta\beta$}

\newcommand{\ITEP}{National Research Center ``Kurchatov Institute'', Kurchatov Complex of Theoretical and Experimental Physics, Moscow, 117218 Russia}
\newcommand{\JINR}{Joint Institute for Nuclear Research, Dubna, 141980 Russia} 
\newcommand{\lbnl}{Nuclear Science Division, Lawrence Berkeley National Laboratory, Berkeley, CA 94720, USA}

\newcommand{\lanl}{Los Alamos National Laboratory, Los Alamos, NM 87545, USA}
\newcommand{\queens}{Department of Physics, Engineering Physics and Astronomy, Queen's University, Kingston, ON K7L 3N6, Canada}
\newcommand{\uw}{Center for Experimental Nuclear Physics and Astrophysics, and Department of Physics, University of Washington, Seattle, WA 98195, USA}
\newcommand{\unc}{Department of Physics and Astronomy, University of North Carolina, Chapel Hill, NC 27514, USA}
\newcommand{\duke}{Department of Physics, Duke University, Durham, NC 27708, USA}
\newcommand{\ncsu}{Department of Physics, North Carolina State University, Raleigh, NC 27695, USA}	
\newcommand{\ornl}{Oak Ridge National Laboratory, Oak Ridge, TN 37830, USA}
\newcommand{\ou}{Research Center for Nuclear Physics, Osaka University, Ibaraki, Osaka 567-0047, Japan}
\newcommand{\pnnl}{Pacific Northwest National Laboratory, Richland, WA 99354, USA}
\newcommand{\ttu}{Tennessee Tech University, Cookeville, TN 38505, USA}
\newcommand{\sdsmt}{South Dakota Mines, Rapid City, SD 57701, USA}
\newcommand{\usc}{Department of Physics and Astronomy, University of South Carolina, Columbia, SC 29208, USA}
\newcommand{\usd}{Department of Physics, University of South Dakota, Vermillion, SD 57069, USA}  
\newcommand{\ut}{Department of Physics and Astronomy, University of Tennessee, Knoxville, TN 37916, USA}
\newcommand{\tunl}{Triangle Universities Nuclear Laboratory, Durham, NC 27708, USA}
\newcommand{\mpi}{Max-Planck-Institut f\"ur Physik, M\"unchen, 80805, Germany}
\newcommand{\tum}{Physik Department and Excellence Cluster Universe, Technische Universit\"at, M\"unchen, 85748 Germany}
\newcommand{\williams}{Physics Department, Williams College, Williamstown, MA 01267, USA}
\newcommand{\ciemat}{Centro de Investigaciones Energ\'eticas, Medioambientales y Tecnol\'ogicas, CIEMAT 28040, Madrid, Spain}
\newcommand{\iu}{IU Center for Exploration of Energy and Matter, Bloomington, IN 47408, USA and Department of Physics, Indiana University, Bloomington, IN 47405, USA}

\graphicspath{{./plot}}

\begin{document}

\ProvideTextCommandDefault{\textonehalf}{${}^1\!/\!{}_2\ $}

\title{Energy Calibration of Germanium Detectors for the \textsc {Majorana Demonstrator}}

\affiliation[a]{\pnnl}
\affiliation[b]{\usc}
\affiliation[c]{\ornl}
\affiliation[d]{\ITEP}
\affiliation[e]{\usd}
\affiliation[f]{\ncsu}
\affiliation[g]{\unc}
\affiliation[h]{\duke}
\affiliation[i]{\tunl}
\affiliation[j]{\uw}
\affiliation[k]{\lbnl}
\affiliation[l]{\sdsmt}
\affiliation[m]{\lanl}
\affiliation[n]{\ciemat}
\affiliation[o]{\ut}
\affiliation[p]{\williams}
\affiliation[q]{\ttu}
\affiliation[r]{\queens}
\affiliation[s]{\mpi}
\affiliation[t]{\tum}
\affiliation[u]{\iu}
\affiliation[w]{\JINR}
\affiliation[x]{\ou}

\author[a]{I.J.~Arnquist,} 
\author[b,c]{F.T.~Avignone~III,}
\author[d]{A.S.~Barabash,}
\author[e]{C.J.~Barton,}	
\author[g,i]{K.H.~Bhimani,}
\author[f,i]{E.~Blalock,}
\author[g,i]{B.~Bos,}
\author[h,i]{M.~Busch,}	
\author[j,1]{M.~Buuck,%
    \note{Present address: SLAC National Accelerator Laboratory, Menlo Park, CA 94025, USA}}
\author[g,i]{T.S.~Caldwell,}	
\author[k]{Y-D.~Chan,}
\author[l]{C.D.~Christofferson,} 
\author[m]{P.-H.~Chu,}
\author[g,i]{M.L.~Clark,}
\author[n]{C.~Cuesta,}
\author[j]{J.A.~Detwiler,}	
\author[o,c]{Yu.~Efremenko,}
\author[x]{H.~Ejiri,}
\author[m]{S.R.~Elliott,}
\author[p]{G.K.~Giovanetti,}  
\author[f,i,c]{M.P.~Green,}  
\author[g,i]{J.~Gruszko,} 
\author[g,i,2, 3]{I.S.~Guinn,%
  \note{Corresponding author}
  \note{Present address: Oak Ridge National Laboratory, Oak Ridge, TN 37830, USA}}
\author[c]{V.E.~Guiseppe,}
\author[g,i]{C.R.~Haufe,}	
\author[g,i]{R.~Henning,}
\author[g,i]{D.~Hervas~Aguilar,} 
\author[a]{E.W.~Hoppe,}
\author[j]{A.~Hostiuc,}
\author[q]{M.F.~Kidd,}	
\author[m,4]{I.~Kim,%
    \note{Present address: Lawrence Livermore National Laboratory, Livermore, CA 94550, USA}}
\author[a]{R.T.~Kouzes,}
\author[b]{T.E.~Lannen~V,}
\author[g,i]{A.~Li,} 
\author[c]{J.M. L\'opez-Casta\~no,} 
\author[g,i,5]{E.~L.~Martin,%
    \note{Present address: Duke University, Durham, NC 27708, USA}}
\author[r]{R.D.~Martin,}	
\author[m]{R.~Massarczyk,}		
\author[m]{S.J.~Meijer,}	
\author[e,6]{T. K. Oli,%
   \note{Present address: Argonne National Laboratory, Lemont, IL 60439, USA}}
\author[e]{L.S.~Paudel,}
\author[u]{W.~Pettus,}	
\author[k]{A.W.P.~Poon,}
\author[c]{D.C.~Radford,}
\author[g,i]{A.L.~Reine,}	
\author[m]{K.~Rielage,}
\author[j]{N.W.~Ruof,}
\author[m]{D.C.~Schaper,}
\author[b]{D.~Tedeschi,}		
\author[c]{R.L.~Varner,}
\author[w]{S.~Vasilyev,}	
\author[g,i,c]{J.F.~Wilkerson,}  
\author[j]{C.~Wiseman,}		
\author[e]{W.~Xu,}
\author[c]{C.-H.~Yu,}
\author[m,7]{B.X.~Zhu,%
    \note{Present address: Jet Propulsion Laboratory, California Institute of Technology, Pasadena, CA 91109, USA.}}
\emailAdd{guinnis@ornl.gov}

\collaboration{\textsc{Majorana} Collaboration}

\date{\today}
\abstract{The \MJD~was a search for neutrinoless double-beta decay (\nonubb) in the $^{76}$Ge isotope.  It was staged at the 4850-foot level of the Sanford Underground Research Facility (SURF) in Lead, SD. The experiment consisted of 58 germanium detectors housed in a low background shield and was calibrated once per week by deploying a $^{228}$Th line source for 1 to 2 hours.  The energy scale calibration determination for the detector array was automated using custom analysis tools.  We describe the offline procedure for calibration of the \Dem~germanium detectors, including the simultaneous fitting of multiple spectral peaks, estimation of energy scale uncertainties, and the automation of the calibration procedure.}


\arxivnumber{}

\maketitle

\section{Introduction}
\label{sec:intro}
Neutrinoless double-beta decay (\nonubb) is a hypothetical, rare nuclear process in which two neutrons in a nucleus simultaneously decay and emit two electrons  but  no  neutrinos. The discovery of this decay would play an important role in understanding the Universe and physics beyond the Standard Model of particle physics. It would demonstrate the Majorana nature of neutrinos~\cite{Majorana:2008,Schechter:1982} and lepton number violation~\cite{Avignone:2007fu,Vergados:2012}, which may provide a mechanism for generating the matter-antimatter asymmetry of the Universe~\cite{Fukugita:1986}. 
Several experiments have placed limits on the half-life of this process beyond $10^{25}$ years~\cite{KamLAND-Zen:2022tow,Agostini:2020,EXO-200:2019rkq,CUORE:2021gpk,Arnquist:2023} and larger experiments are being developed with half-life discovery potentials of up to $10^{28}$ years~\cite{LEGEND:2021bnm,nEXO:2021ujk,CUPID:2019imh}. 

Some of these recent and planned experiments consist of arrays of solid-state detectors with a total mass of $10$ to $1000$~kg, including the~\MJD, GERmanium Detector Array (GERDA)~\cite{Agostini:2020}, Large Enriched Germanium Experiment for Neutrinoless $\beta\beta$-Decay (LEGEND)~\cite{LEGEND:2021bnm} and Cryogenic Underground Laboratory for Rare Events (CUORE)~\cite{CUORE:2019yfd,CUORE:2021gpk}. With typical detectors each having a mass of order 1~kg for these experiments, the accurate calibration of a large number of detectors is an important issue for timely analysis of data.  Since \nonubb~experiments consist of a search for a relatively narrow peak in an energy spectrum, the corresponding energy uncertainty is a key experimental parameter that must also be estimated. 

The \Dem~experiment was composed of 58 high-purity germanium (HPGe) detectors, using the Broad Energy Germanium (BEGe), P-type Point Contact (PPC)~\cite{Luke:1989,Barbeau:2007}, and Inverted Coaxial Point Contact (ICPC)~\cite{Cooper:2011} detector geometries. These detectors were divided between two independent cryostats, which were built and deployed sequentially. 
 
The first cryostat (Module~1) operated from Jun. 2015 to March 2021 with 20 PPC germanium detectors (16.8~kg) enriched to 88\% in $^{76}$Ge~\cite{Abgrall:2017acl} and 9 natural BEGe detectors (5.6~kg). 
The second cryostat (Module~2) operated from August 2016 to November 2019 with 15 enriched PPC detectors (12.9~kg) and 14 natural BEGe detectors (8.8~kg).  Module~2 was upgraded and operated from September 2020 to March 2021 with 4 enriched ICPC detectors (6.7~kg), 9 enriched PPC detectors (7.4~kg), and 14 natural BEGe detectors (8.8~kg). 

The cryostats and detector support structure were constructed from ultra-low background materials, including underground electroformed copper and polytetrafluoroethylene. These were installed within a graded, low--back\-ground shield~\cite{Abgrall:2014,Abgrall:2016cct,Alvis:2019sil}. 
The $^{228}$Th line source deployment system integrated into the cryostat modules, used in the calibration, was thoroughly described in Ref.~\cite{Abgrall:2017gpr}. 

Charged particles and photons that deposit energy in a Ge detector create charge clouds that 
are collected by a charge sensitive amplifier circuit consisting of a low-mass front end (LMFE) board connected to the point contact of each detector.  In the \Dem, the LMFE output was connected to a preamplifier card placed outside of the shielding. 
Each detector signal was divided between two preamplifier channels with different gain factors, $\times$10 for the high gain and $\times$3 for the low gain.  The complete readout system was described in Ref.~\cite{Abgrall:2022jinst}.
The digitization process, threshold estimation, corrections for ADC non-linearities \cite{Abgrall:2020jto} and charge trapping in the crystal were described in Ref.~\cite{Arnquist:2023ct}.  From these corrected waveforms, after trapezoidal filtering, we estimated the uncalibrated energy and other energy estimators.

The \Dem~data were divided into several separate datasets based on significant modification of experimental configuration during construction and commissioning, such as installation of copper shield, installation of new modules, integration of data acquisition (DAQ) system, etc.  The details of our dataset organization have been described in Ref.~\cite{Alvis:2019sil}. 
The \Dem~was calibrated once per week for 60 to 120 minutes per module in order to minimize the systematic error from calibration drift and maximize the physics data taking period, as well as to monitor the detector stability. 
The length of the calibration period increased over the life of the experiment to compensate for the decrease in the $^{228}$Th source strength due to its 1.9~year half-life.  Approximately every 2~months, the \Dem~also recorded   calibrations of up to 17 hours for each module, referred to as long calibration runs.   These runs provided the statistics necessary for tuning the pulse-shape discrimination parameters used to reject multisite events~\cite{AvsE:2019} and surface alpha events~\cite{Gruszko:2022}. 
We calibrated the energy spectrum by matching the true energy of known $\gamma$-rays emitted in the $^{228}$Th decay chain with the positions of peaks in the uncalibrated energy spectrum. 

This paper describes the determination of the energy calibration in the \MJD\  search for \nonubb~\cite{Aalseth:2017,Alvis:2019sil,Arnquist:2023}.   The excellent energy resolution of the large Ge semiconductor detectors in this experiment was critical to the search for \nonubb .  Instrumental fluctuations over time can lead to small changes in the calibration.  If these changes are tracked through regular calibration measurements,  the experiment can maintain excellent resolution over several years of operation.  The overall energy resolution (full width at half maximum, FWHM) that we have achieved is $2.52\pm0.08$ keV at 2039 keV~\cite{Arnquist:2023}, the lowest among the current \nonubb~experiments.

We describe the offline calibration procedure and the analysis tools used to accurately estimate the energy of events.  
Section~\ref{sec:peakshape} describes the procedure for fitting peaks in the measured calibration spectra, which is the first step to obtain the uncalibrated energy values of the gamma peaks. 
Section~\ref{sec:calibration} describes the procedure to calibrate the energy scale of the \MJD\ and Section~\ref{sec:uncertainties} describes the estimation of the associated systematic uncertainties. 
 
This procedure can be efficiently used for the next generation experiments with hundreds of germanium detectors, 
such as the Large Enriched Germanium Experiment for Neutrinoless $\beta\beta$ Decay (LEGEND)~\cite{LEGEND:2021bnm}.

\section{Single- and Multi-Peak Shape Fitting}
\label{sec:peakshape}

An energy calibration can be performed by measuring moments of spectral peaks or by fitting a Gaussian distribution to the peaks.  The principal ingredient is a Gaussian function
\begin{linenomath*}
\begin{align}
\mathrm{G}(E) =& \frac{A(1-f_{LE}-f_{HE})}{\sqrt{2\pi}\sigma}\exp\left(-\frac{(E-\mu)^2}{2\sigma^2}\right) \label{eq:Gaus}
\end{align}
\end{linenomath*}
where, 
\begin{itemize}
   \item $\mu$ = mean of Gaussian function,
   \item $\sigma$ = standard deviation of Gaussian function, 
   \item $A$ = peak area; total number of counts in the Gaussian and tail functions, and
   \item $(1-f_{LE}-f_{HE})$ is the fraction of the total peak area in the Gaussian function, where $f_{LE}$ and $f_{HE}$ are the fraction of the total area taken up by the low energy (LE) and high energy (HE) tails subject to the constraint that
   \begin{align*}
       0  \leq & f_{LE} +f_{HE}  \leq 1
   \end{align*} 
   as defined in equation~\ref{eq:Tail}~and following.
\end{itemize}

HPGe detector peaks often have features such as low energy tails and steps underneath the peaks that can introduce biases in calibration parameters obtained using a simple Gaussian function, degrade energy resolution by misaligning peak shapes or result in inaccurate estimates of the detection efficiency for a chosen region of interest. 
The tail functions are represented by the exponentially modified Gaussian function, in which an exponential distribution with tail length $\gamma_\alpha$ is convolved with a Gaussian using the same parameters as the Gaussian peak shape component, such as
\begin{linenomath*}
\begin{align}
\mathrm{T}_{\alpha}(E) =& \frac{Af_\alpha}{2\gamma_\alpha}\exp\left(\frac{\sigma^2}{2\gamma_\alpha^2}\pm\frac{E-\mu}{\gamma_\alpha}\right) 
\mathrm{erfc}\left(\frac{\sigma}{\sqrt{2}\gamma_\alpha}\pm\frac{E-\mu}{\sqrt{2}\sigma}\right)\label{eq:Tail}
\end{align}
\end{linenomath*}
where,
\begin{itemize}
\item $\alpha$=LE(+),~{HE}(-), the signs of which correspond to the choice of $\pm$ above, and
\item $\gamma_{LE}$ or $\gamma_{HE}$ = decay constant of the LE/HE tail exponential.
\end{itemize}

Reduction in charge collection because of trapping in the detector bulk can produce a low energy tail.
Imperfect deconvolution of the electronics response function by the pole-zero correction can produce either a high or low energy tail. 
The low energy tail was reduced, and the high energy tail made negligible, through optimizations in energy estimation described in~\cite{Alvis:2019sil}. 

A step background is produced by low angle scattering of $\gamma$ rays resulting in small energy loss prior to full absorption in the detector, and by degradation in charge collection in the transition region between the surface dead layers and the detector bulk,
\begin{linenomath*}
\begin{align}
\mathrm{S}(E) =& \frac{AH_s}{2}\mathrm{erfc}\left(\frac{E-\mu}{\sqrt{2}\sigma}\right) \label{eq:Step}
\end{align}
\end{linenomath*}
in which
\begin{itemize}
  \item $H_{s}$ = height of step background as a fraction of the peak area.
\end{itemize}

Gamma peaks in Ge detectors are commonly fit using a combination of these functions.   
The model used to describe the \Dem~calibration gamma peaks was composed of the analytic functions just discussed in 
equations~\ref{eq:Gaus},~\ref{eq:Tail}, and~\ref{eq:Step}~that have been used to fit HPGe data~\cite{Longoria:1990,Hammed:1993,Kanisch:2017}.
\begin{linenomath*}
  \begin{align}
    \mathrm{PS}(E) =& \mathrm{G}(E) + \mathrm{T_{LE}}(E) + \mathrm{T_{HE}}(E) + \mathrm{S}(E) \label{eq:PS}
  \end{align}
\end{linenomath*}

When modeling real data, a quadratic background function ($\mathrm{BG}(E)$) was added to the peak shape function:
\begin{linenomath*}
  \begin{align} 
     \mathrm{BG}(E)=& qP_2(E-E_{cen})+mP_1(E-E_{cen})+b
     \label{eq:BG}
  \end{align}
\end{linenomath*}
where $E_{cen}$ is the center of the energy domain this function is being fit on, and $P_1$ and $P_2$ are Legendre polynomials. The parameters that are allowed to float while fitting are $b$, a flat background component, $m$, the coefficient of the linear component, and $q$, the coefficient of the quadratic component.

The full background model was quadratic; however, we often used a linear background (fixing $q$ to 0) or flat background 
(fixing $m$ and $q$ to 0).  The choice of background model was based on the amount of statistics in the data used for a fit; 
for example, around the 2615~keV $^{208}$Tl peak in the calibration spectrum, there are usually too few counts to use a quadratic or linear model. $\mathrm{BG}(E)$ was defined only in a limited range of energies centered around $E_{cen}$, selected to be large enough that the background function contributions were separable from peak contributions, but small enough that a quadratic model was a valid approximation of the data. Thus, we modeled background around a single peak by summing equations \ref{eq:PS} and \ref{eq:BG}, resulting in a model with up to 11~parameters. 

\begin{figure}[t]
    \includegraphics[width=\columnwidth]{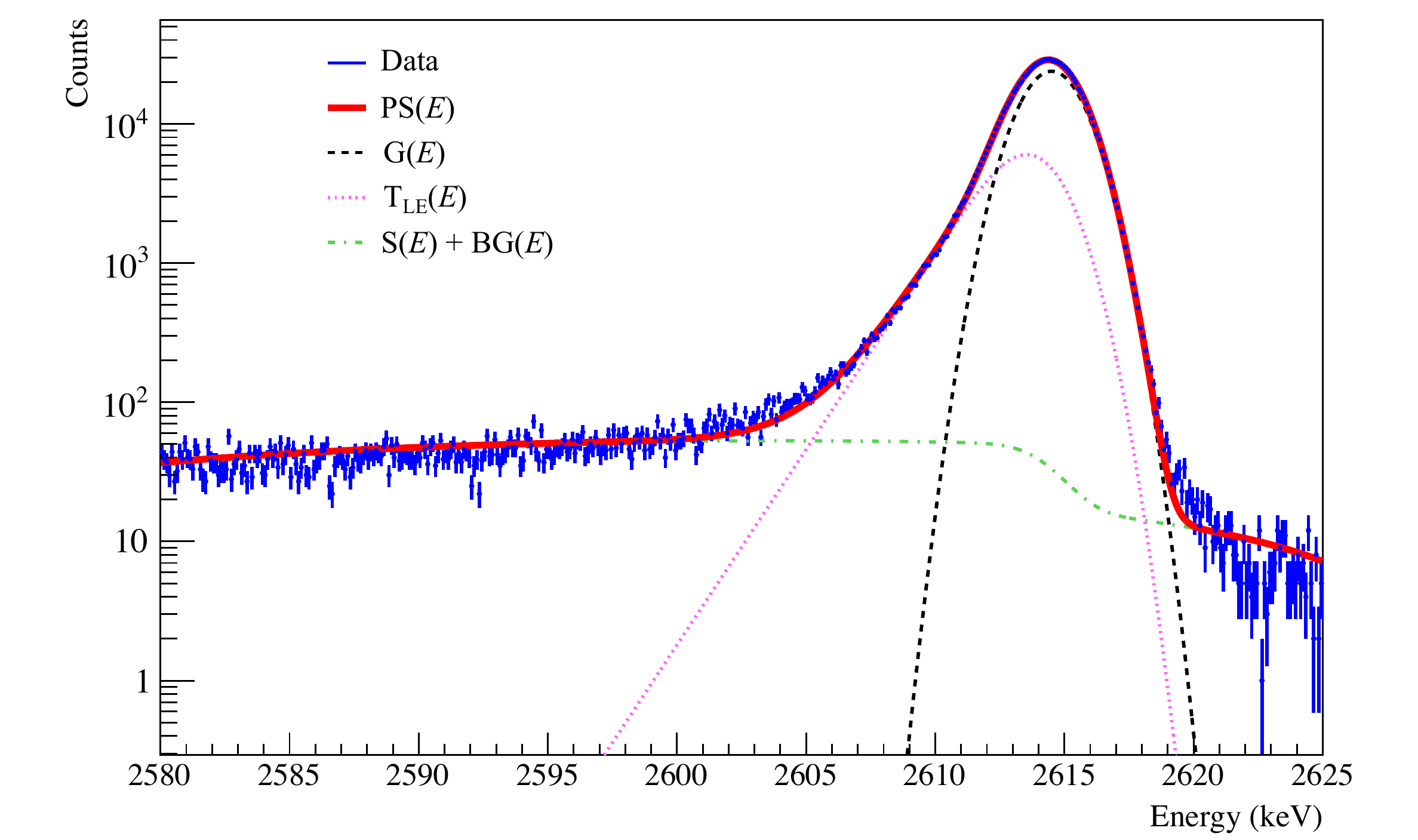}
    \caption{The peak shape model fitted using "Algorithm 1" applied to the 2615~keV $^{208}$Tl peak, combining data from all detectors in all datasets in Ref.~\cite{Alvis:2019sil}.  The peak shape function (solid red) included Gaussian (short-dash black), low energy tail (dot magenta) and step functions (dash-dot green), as well as the quadratic background.  The FWHM of the peak is 2.95~keV.   By fitting multiple peaks in the calibration spectrum, we can calculate a FWHM of 2.5 keV at the 2039 keV Q-value for \nonubb.  The exclusion of a high-energy tail was automatically selected by comparing AIC (Akaike Information Criterion) for various models.  The figure was modified from~\cite{Alvis:2019sil}.}
  \label{fig:peakshape}
\end{figure}

Our first approach (Algorithm 1) to fitting energy peaks in a histogram was to fit each peak separately.  In fitting a single peak, we computed the fitting function and numerically evaluated properties of the model peak such as the centroid and the FWHM.   A Poisson negative log-likelihood (nLL) compared this model to a histogram of the energy for the purpose of fitting; in order to improve the speed and reliability of this process, we analytically computed the parameter gradients of the fitting function. 
The MINUIT fitting package MIGRAD algorithm~\cite{James:1975,ROOT} was used to perform maximum likelihood fits. 
We utilized an additional fitting algorithm to perform multiple fits with different sets of parameters enabled, using the Akaike Information Criterion (AIC), which penalized the addition of parameters to a model in order to avoid overfitting~\cite{Akaike:1974}. 
This algorithm first performed fits, with the flat and step background components enabled, of peaks with different functions in the following sequence: \textbf{1.}~$\mathrm{G}(E)$, \textbf{2.}~$\mathrm{T_{LE}}(E)$, \textbf{3.}~$\mathrm{T_{HE}}(E)$, \textbf{4.}~$\mathrm{G}(E)+\mathrm{T_{LE}}(E)$, \textbf{5.}~$\mathrm{G}(E)+\mathrm{T_{HE}}(E)$, and then \textbf{6.}~$\mathrm{G}(E)+\mathrm{T_{LE}}(E)+\mathrm{T_{HE}}(E)$. 
After rejecting fits that failed to converge, the best fit according to AIC was chosen and fit with a flat, linear and/or quadratic background, and the best overall fit among these was used. 
Figure~\ref{fig:peakshape} shows an example of this model with a quadratic background fit to data around a 2615~keV $^{208}$Tl combined peak from all detectors and all datasets. 

This fitting algorithm worked well for many energy calibration purposes.  However, fitting single peaks in the energy calibration spectra had several limitations. 
First, calibration data taken using a $^{228}$Th source had several peaks that are too close together to use a single peak model. For example, one of the most prominent peaks, at 239~keV, was right next to another at 240~keV. 
Second, for peaks with a poor signal to background ratio, correlated errors between different parts of the peak, such as the step and LE~tail, can result in inaccurate fits. 
Occasionally, the model parameters determined using this sequential algorithm based on the AIC varied significantly because of statistical fluctuations in the peak shape. This would cause the systematics of detector response versus energy to be inconsistent. While single peak fits were useful in some circumstances, a simultaneous fit to multiple peaks helps reduce these difficulties.

\begin{figure*}[t]
\centering
\includegraphics[width=\textwidth]{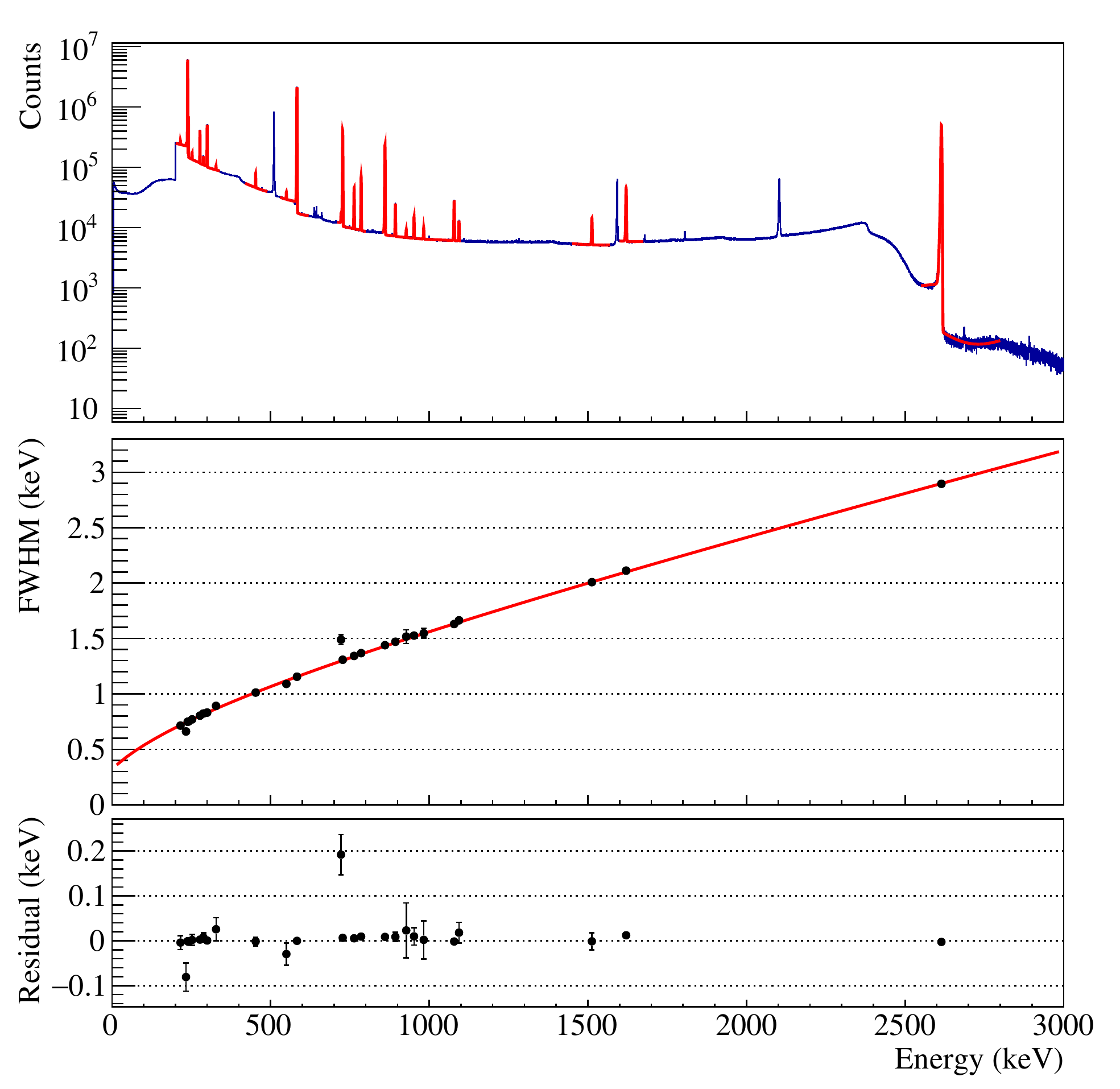}
\caption{At the top, combined energy spectrum including all data recorded with all detectors during $^{228}$Th calibration runs from one of our data subsets containing 8 months of data (top panel). The Algorithm 2 multi-peak fitter was used to fit a collection of 24 peaks, and the FWHM energy resolution computed as a function of energy (middle panel). This function was used to compute the FWHM energy resolution for each peak, and at the 2039~keV Q-value of \nonubb~in $^{76}$Ge.  A second fit was performed floating the peak width parameters, and used to compute the residuals and uncertainties for each peak (bottom panel). These residual measurements were used to perform studies of systematic errors in the \Dem's energy estimation.  Points with large residuals were mostly very small peaks on the shoulder of larger peaks.}
\label{fig:multipeakfwhm}
\end{figure*}

To improve the reliability of our energy calibration process, we developed an algorithm (Algorithm 2) in which multiple peaks were simultaneously fit using a multilevel model, meaning each peak's model parameters are calculated from a smaller set of hyper-parameters that describe each model parameter as a function of the peak energy.  
This involves dividing the calibration spectrum into separate energy regions, $j$, containing one or more peaks, $i$, each of which corresponds to a gamma ray of known energy, $E_i$, from the calibration source, 
\begin{linenomath*}
\begin{align}
    R_j(E) = \sum_i \mathrm{PS}_i(E) + \mathrm{BG}_j(E)
    \label{eq:region}
\end{align}
\end{linenomath*}
Each region had an independent set of three background parameters, and
each peak shape function $\mathrm{PS}_i(E)$ had its own set of peak shape parameters ($A_i$, $\mu_i$, $\sigma_i$, $f_{\mathrm{LE/HE},i}$, $\gamma_{\mathrm{LE/HE},i}$ and $H_{s,i}$). These were either calculated from a parametric function of the physical peak energy and corresponding hyper-parameters, or they were floated independently for each peak. For the sake of clarity, we use bold-face for hyper-parameters. 
The peak shape parameters typically depended on energy of the gamma ray, E$_i$, as follows: 
\begin{linenomath*}
\begin{align}
A_i =& \boldsymbol{A_i}, \label{eq:multifitmodelstart}\\
\mu_i =& \boldsymbol{\mu_0} + \boldsymbol{\mu_1}E_i\label{eq:multifitmu}\\
\sigma_i =& \sqrt{\boldsymbol{\sigma_0}^2 + \boldsymbol{\sigma_1}^2E_i + \boldsymbol{\sigma_2}^2E_i^2}\label{eq:multifitsigma}\\
\gamma_{\mathrm{LE},i} =& \boldsymbol{\gamma_{\mathrm{LE},0}} + \boldsymbol{\gamma_{\mathrm{LE},1}}E_i\label{eq:multifitgammale}\\
f_{\mathrm{LE},i} =& \boldsymbol{f_{\mathrm{LE},0}}\\
\gamma_{\mathrm{HE},i} =& \boldsymbol{\gamma_{\mathrm{HE},0}} + \boldsymbol{\gamma_{\mathrm{HE},1}}E_i\label{eq:multifitgammahe}\\
f_{\mathrm{HE},i} =& \boldsymbol{f_{\mathrm{HE},0}}\\
H_{s,i}=&\frac{\boldsymbol{H_{s,0}}}{E_i^2} + \boldsymbol{H_{s,1}}E_i^{-0.88} \label{eq:multifitmodelend}
\end{align}
\end{linenomath*}
While the model used was empirical, the origin of the hyper-parameters can be understood from physical sources. $\boldsymbol{\mu_0}$ and $\boldsymbol{\mu_1}$ represent the energy scale calibration constants for the detectors. $\boldsymbol{\sigma_0}$ and $\boldsymbol{\gamma_{\mathrm{LE/HE},0}}$ arise from electronic noise, $\boldsymbol{\sigma_1}$ from the detector Fano factor, and $\boldsymbol{\sigma_2}$ and $\boldsymbol{\gamma_{\mathrm{LE/HE},1}}$ from sources of broadening such as charge trapping.  The first component of the step height model $\boldsymbol{H_{s,0}}$ arises from small angle scattering of gammas before absorption by detectors. The $E^2$ dependence of this can be analytically derived~\cite{Oberer:2011}.  The second component of the step height $\boldsymbol{H_{s,1}}$ arises from the detector dead layers. The $E^{-0.88}$ proportionality was derived by varying the transition dead layer profile in Monte Carlo simulations. We found that a power of $-0.88$ fit the steps in the Monte Carlo data well, with very little correlation with transition dead layer parameters~\cite{Guinn:2019}. In addition, this power law fit the steps found in calibration data for both the \Dem~PPC and BEGe germanium detectors. Figure~\ref{fig:multipeakfwhm} shows how the functions described relate to the FWHM peak width in a calibration energy spectrum, using the above hyper-parameters as an example.

The parameters were fit to data using a maximum likelihood fit. The log-likelihood was the sum of the log-likelihood contributions from each of the ~$j$ regions, computed using a Poisson likelihood function comparing equation~\ref{eq:region} to bin contents in region~$j$. 
Due to the large number of fit parameters and the high correlation between many of these parameters, a successful fit was highly sensitive to the initial parameter guess.  For a modular array with a large number of detectors, manually tuning these initial parameters was not feasible. To enable successful convergence of MIGRAD fits 
starting from fixed initial parameter guesses, a Hamiltonian Monte Carlo (HMC)~\cite{Duane:1987,Neal:2011,Betancourt:2017} algorithm was first used. 
HMC is a Markov Chain Monte Carlo technique that generates steps using the leapfrog algorithm, which utilizes gradient information to increase the probability of accepting any given step even with relatively long step lengths. To further improve the rate of convergence, a variant on Riemann Manifold HMC (RMHMC)~\cite{Girolami:2011} was used, varying the metric based on the Fisher information matrix. This variant does not properly evolve the metric, and does not asymptotically converge to the correct distribution; however, it dramatically increases the rate of convergence towards the mode of the distribution and suffices for burn-in steps. In addition, the length of each leapfrog step is adjusted between each step, depending on whether the step is accepted or rejected. Typically, 200 HMC burn-in steps with 50 leapfrog steps were used to obtain a successful fit compared to tens of thousands required for ordinary MCMC. The most likely parameters sampled were then used as inputs for a gradient minimization, performed using the MIGRAD algorithm implemented in the MINUIT package~\cite{James:1975,ROOT}. 

\begin{figure*}[t]
\centering
\includegraphics[width=\textwidth]{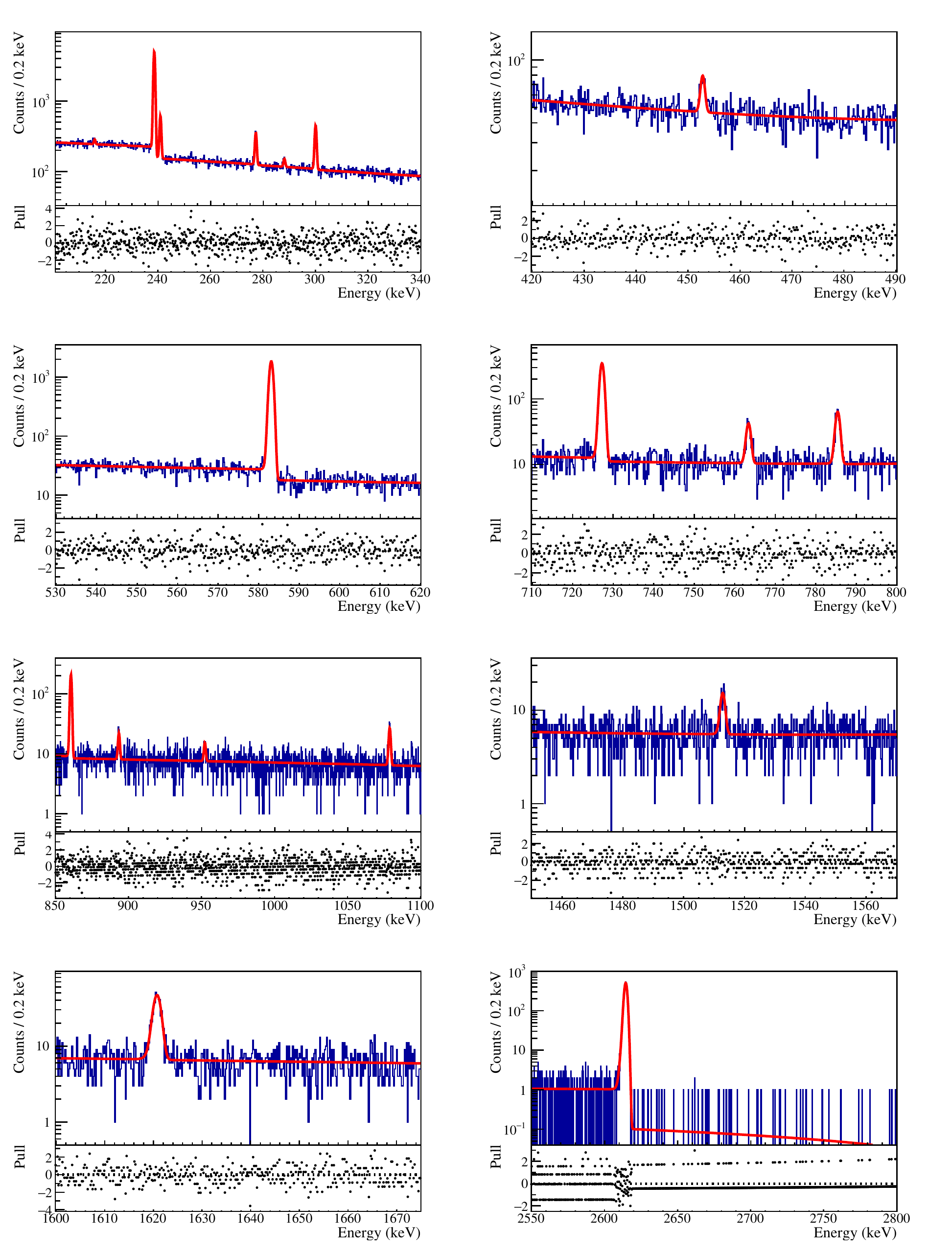}
\caption{An example of an Algorithm 2, multi-peak fit of eighteen peaks from a single detector during a 41-hour calibration run. This fit used the model described by equations~\ref{eq:multifitmodelstart}-\ref{eq:multifitmodelend}, with $f_{HE,0}$ fixed to 0. }\label{fig:multipeakspectrum}
\end{figure*}
A multi-peak fit following this procedure enabled fitting of lower amplitude peaks that could not be reliably fit individually. 
The statistics from a weekly 1.5 hour calibration period was sufficient to fit 8 peaks between 200~keV and 2615~keV in the $^{228}$Th decay chain. 
Combining multiple calibration runs or spectra from multiple detectors provided the statistics to fit additional peaks simultaneously. 
For example, figure~\ref{fig:multipeakspectrum} shows the result of a simultaneous fit of eighteen peaks from a monthly 41-hour calibration period. 
In order to identify and measure systematic errors, the fitter can vary the model used in performing these simultaneous fits; specifically, a parameter can be made independent of hyper-parameters in order to explore the residuals between the fitted hyper-parameters and the individually fitted parameters. Figure~\ref{fig:multipeakfwhm} shows an example of this process, applied to the FWHM energy resolution. 

We computed the FWHM and centroid of peaks from the peak shape parameters or hyper-parameters. The hyper-parameters were also used to compute analysis parameters for $0\nu\beta\beta$ and other peak searches, such as the proportion of counts in a peak that occur within a limited-width region of interest, and the optimal region of interest for measuring a peak at a given energy with a specified background level. The statistical uncertainties of these parameters were computed using a multi-variable error propagation with the full covariance matrix of a fit.

\section{Calibration Procedures and Analysis Method for Weekly and Combined Calibration Data}
\label{sec:calibration}

\begin{figure}
    \centering
    \includegraphics[width=\columnwidth]{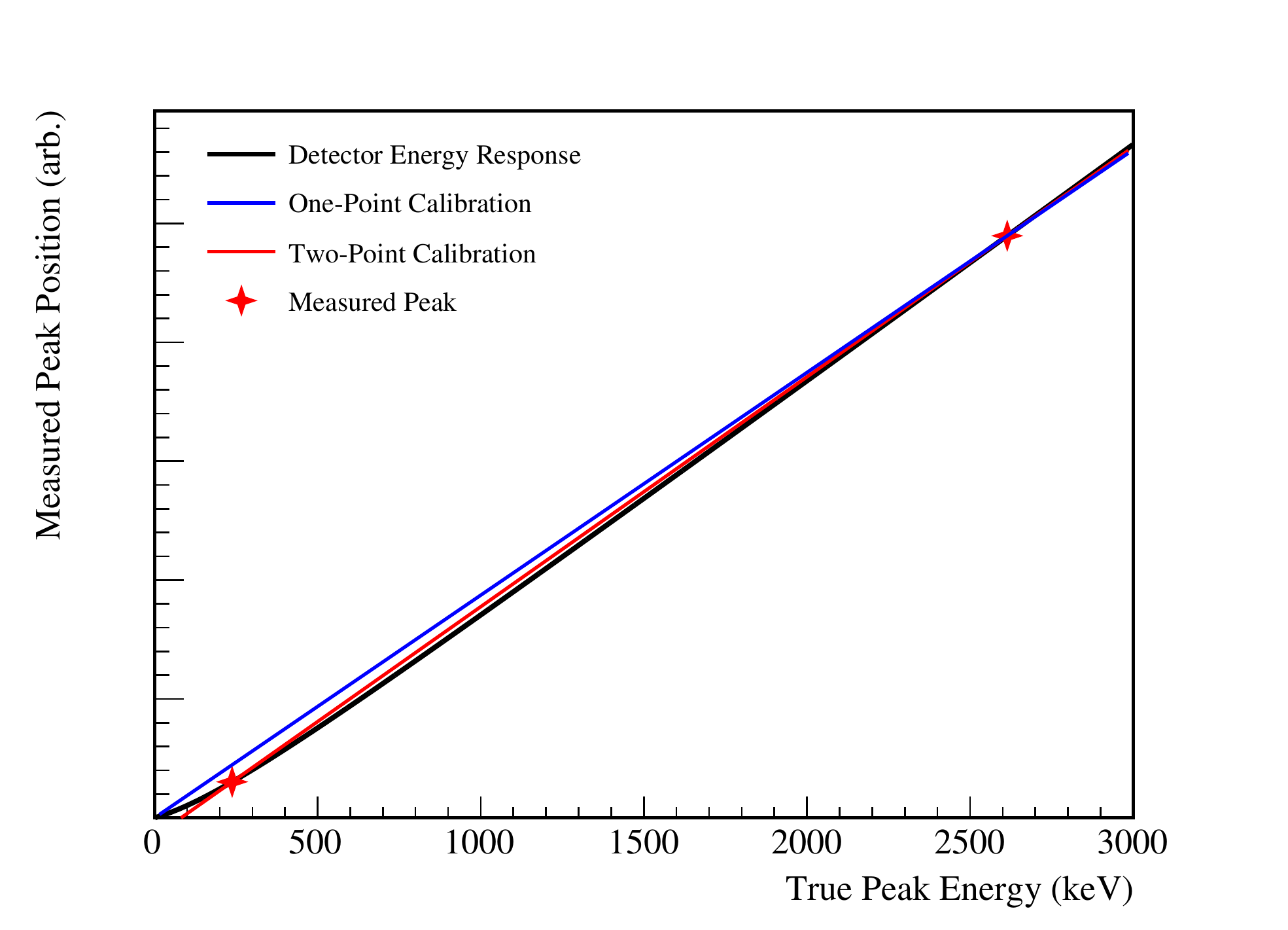}
    \caption{A simple toy model illustrating the differences between the first and second stages of calibration for each procedure in section~\ref{sec:calibration}.  The energy response for our uncalibrated energy estimator is represented with a hyperbolic curve (black), and we label the measured position of the two most prominent calibrations peaks at 239 keV and 2615 keV (red crosses). The detector has a global nonlinearity that is greatest below 200 keV, the size of which we exaggerate by a factor of about 200 for illustration purposes. For all three approaches, the first stage calibration (blue) uses low statistics and relies solely on the 2615 keV peak and the E=0 intercept; this is most accurate near these energy regions. We approximate our second stage calibration using a two-point calibration (red) using the 239 keV and 2615 keV peaks; this is more accurate between these energies, but gains a non-zero intercept at E=0.} 
    \label{fig:cal_cartoon}
\end{figure}

In this section, we describe three automatic calibration procedures that were used in the \nonubb~search with the \Dem~reported in Refs.~\cite{Aalseth:2017},~\cite{Alvis:2019sil}, and \cite{Arnquist:2023} and the calculation of the corresponding energy uncertainties.  
Each procedure has two stages.   First, calibrations are performed weekly based on a linear fit dominated by the $^{208}$Tl 2615~keV peak and $E=0$ intercept. 
Second, a correction is performed using data from many calibrations with multiple $^{228}$Th chain peaks ranging from 238~keV to 2615~keV. This correction improves energy estimation in this range, which includes the 2039~keV region of interest, due to nonlinearities in the energy response at energies $<200$~keV. Figure~\ref{fig:cal_cartoon} demonstrates, using an exaggerated toy model, the utility of this correction.

\subsection{Approach 1}
The first stage of the calibration procedure followed in Ref.~\cite{Aalseth:2017} was to perform individual fits of four prominent gamma peaks in the calibration spectrum. The first step located the prominent 238.6, 583.2, 727.3, and 2614.5~keV gamma peaks of the $^{228}$Th decay chain in the uncalibrated energy spectrum.  Since the 2615~keV peak was isolated, very prominent, and usually located in a fixed range, an initial estimate of the 2615~keV peak was reliably taken as the maximum count in the spectrum in the expected ADC code range. We estimated the location of other peaks in the uncalibrated energy spectrum based on a linear scaling between the energy and ADC codes of the 2615~keV peak.  
We then individually fitted these four gamma peaks using Algorithm 1 described in section~\ref{sec:peakshape}. From the fit results, we calculated the peak centroid and centroid uncertainty to determine the peak locations in the spectrum in ADC codes.

\begin{figure}[t]
\centering
\includegraphics[width=\columnwidth]{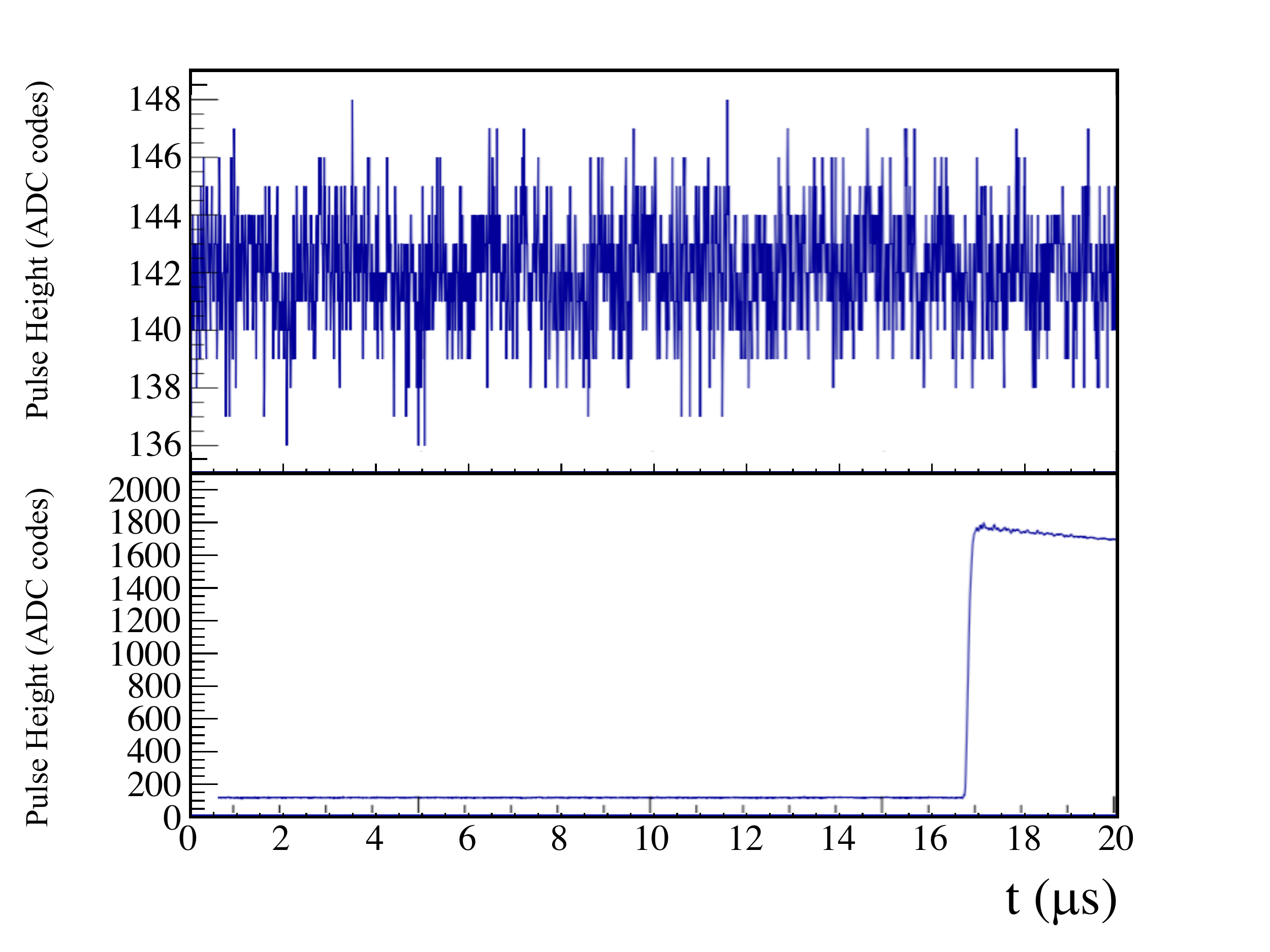}
\caption{An example of a E=0 point waveform. (top) Artificially triggered by the data-acquisition system; (bottom) Pulser data shifting the rising edge to the very end of the digitizer time window.}
\label{fig:ezero}
\end{figure}

An additional peak at E=0 was found by applying our usual trapezoidal filter~\cite{Alvis:2019sil} to a signal that deposits no energy (E) in the detector. 
Events in the E=0 peak for each detector of each data set were obtained by either artificially triggering the data-acquisition system or by using pulser data with the rising edge shifted to the very end of the digitizer sample window as shown in figure~\ref{fig:ezero}. 
As expected, the raw output energy was very close to zero, between -0.4 keV and 0.4 keV for working detectors.
This peak was evaluated using Algorithm 1, and its position found by computing the centroid and centroid uncertainty of the peak.

Using the E=0 peak and the other four gamma peak positions obtained by the single peak fitter, we performed a weighted fit to a linear function from the known gamma energies to the peak positions. From the fitted parameters, we computed linear calibration coefficients for the energy scale $a_1$ and the offset $a_0$, defined as 
\begin{linenomath*}
\begin{align}
 E_{\text{cal1}} = a_0 + a_1 E_{\text{uncal}}.
 \label{eq:linear}
\end{align}
\end{linenomath*}
Due to the much higher statistics of the E=0 and 2615~keV peaks they dominate the fit, producing a result similar to the "One-Point Calibration" in figure~\ref{fig:cal_cartoon}.  The parameters $a_0$ and $a_1$ were uploaded to a database and were applied to all physics data until the next calibration period to convert the uncalibrated energy in ADC codes, $E_{\text{uncal}}$, to the calibrated energy in keV, $E_{\text{cal1}}$. 

Using the E=0 point improved the accuracy of the calibration at low energy by constraining the offset of the linear fit. However, an energy nonlinearity existed because there was a small bias in estimating the start time $t_0$ of the pulse rise, that became more significant at low energy due to the decreased signal-to-noise ratio~\cite{Arnquist:2023ct}. This nonlinearity resulted in a shift in low energy ($<$ 1 MeV) calibrated peak positions by up to 0.15 keV relative to their true energies. In order to correct this shift when we combined all calibrated spectra from a dataset as described in the next paragraph, we recalibrated the energy without the E=0 point to the energy spectrum of $E_{\text{cal1}}$, using more peaks in the spectrum. This energy correction was
\begin{linenomath*}
\begin{align}
\label{eq:cal_corr}
    E_{\text{cal2}} = a^\prime_0 +a^\prime_1 E_{\text{cal1}}
\end{align} 
\end{linenomath*}
with the energy scale correction $a^\prime_1$ and the energy offset $a^\prime_0$.  This corrected energy $E_{\text{cal2}}$ was used for the analyses of \Dem~data.   However,  as in the "Two-Point Calibration" in figure~\ref{fig:cal_cartoon}, at $<1$~MeV, the corrected calibration incorrectly shifted energies due to a non-zero $a^\prime_0$, so the low energy analysis~\cite{Vorren:2016} used $E_{\text{cal1}}$ without the second calibration.

To determine the recalibration, E$_{\text{cal2}}$, we used accumulated calibration data including all available weekly calibration periods and long calibration periods in the data set for each detector.  Algorithm 2 was applied to simultaneously fit eight prominent peaks in the $^{228}$Th decay chain at 238.6, 241.0, 277.4, 300.1, 583.2, 727.3, 860.6, and 2614.5~keV.  
The $\mu$ parameter was floated independently for each peak.  Initial parameters for the fit were area and location of the peaks, the remaining hyper-parameters, and the background.  Since we were fitting to a previously calibrated spectrum, the initial peak positions were estimated as the true gamma energies. The other fit parameters were initialized to a pre-constructed template and the fits performed using RMHMC with the \textsc{Minuit} final optimization.  The best fit calibration parameters, covariance matrix and fitting parameters from the multi-peak fitter were stored in the database~\cite{mjapdb:2022} for each calibration period. 

The energy calibration procedure Approach~1, used to produce the \Dem~results in Ref.~\cite{Aalseth:2017}, was robust; however, it required an additional manual check and occasional correction of fitting errors because of the relatively low statistics of the regular calibrations. During most weekly calibrations, at least one detection channel required manual intervention to achieve optimal energy stability.

\subsection{Approach 2}
\begin{figure}[hb]
\includegraphics[width=\columnwidth]{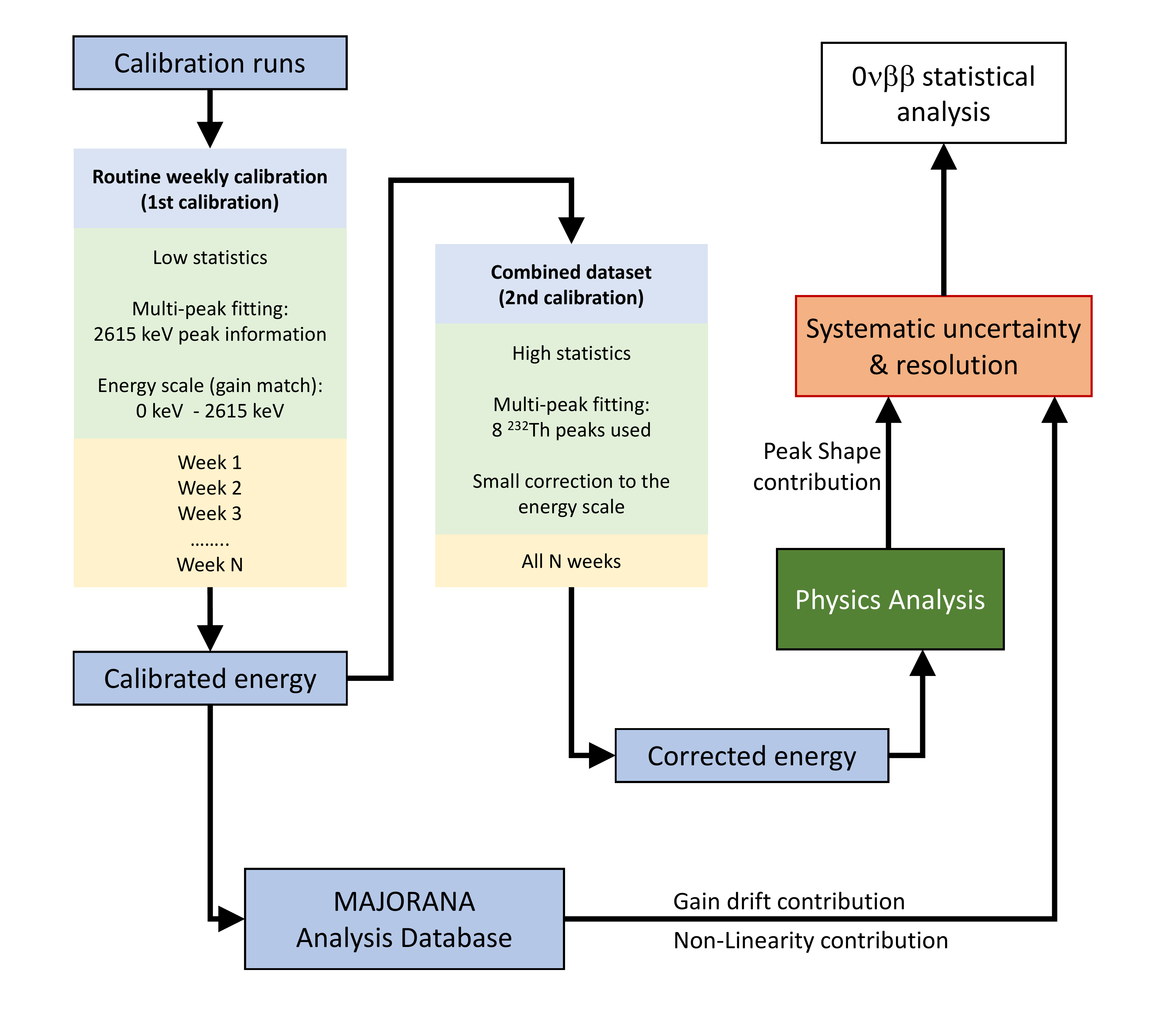}
\caption{Flowchart depicting the procedure for energy calibration Approach 2, the gain matching method.}
\label{fig:flowchart}
\end{figure}

To overcome the weaknesses in Approach~1 and to minimize human involvement in our analysis, we developed the automated procedure for calibration reported in Ref.~\cite{Alvis:2019sil}.  Similar to Approach~1, we applied a two-step process, shown schematically in figure~\ref{fig:flowchart}. 
First, we extracted the peak position,  $\mu_{2615}$, of the 2615~keV peak and calculate
\begin{linenomath*}
\begin{align}
    & E_{\text{cal1}} = a_1 E_{\text{uncal}}\label{eq:gainmatch}\\\nonumber
    & a_1=\frac{2614.533}{\mu_{2615}}
\end{align} 
\end{linenomath*}
equivalent to a linear fit with no offset. This method does not require an E=0 peak, and implicitly assumes that these events will measure 0~ADC codes with a result comparable to "One-Point Calibration" in figure~\ref{fig:cal_cartoon}.  The position of the 2615~keV peak is extracted using the multi-peak fitter, with the same set of peaks as the second stage fit from Approach~1.
While the positions of the other peaks are ignored for this fit, they provide valuable information about the peak shape that improves the stability of the fit at the low statistics of our weekly calibration runs.
We call this technique "gain matching", and it provides a high degree of stability in fit results with minimal human intervention.
The value of $a_1$ is stored in the database and used to calibrate energies for all physics data until the next calibration period.

Finally, we perform a second fit to a single histogram that accumulates the calibration data for an entire dataset.  This fit is done with no assumption about E=0, applying the same procedure as in Approach~1 in order to produce a corrected calibration that is more accurate from 238~keV to 2615~keV, as in the "Two-Point Calibration" of figure~\ref{fig:cal_cartoon}. Following Approach~2, we rarely required manual intervention, and achieved a similar energy stability and resolution to Approach~1.

\subsection{Approach 3}
A final approach to energy calibration was applied to the \Dem~results in Ref.~\cite{Arnquist:2023}.
For this procedure, we use an energy estimator that corrects for the energy non-linearities at low energies.  These result from a bias in the leading-edge filter used for charge trapping correction~\cite{Arnquist:2023ct}.
Once again, we apply a two-step procedure, with the first stage following the same gain-matching approach used in Approach~2.
For the second stage of calibration, we apply the multi-peak fitter to the accumulated calibration data for a full dataset, using the same peaks as in the other approaches. However, unlike before, we now fit to a quadratic energy response
\begin{linenomath*}
\begin{align}
    E_{\text{cal2}} = a^\prime_1 E_{\text{cal1}} + a^\prime_2 E_{\text{cal1}}^2
    \label{eq:finalcal}
\end{align}
\end{linenomath*}
with linear and quadratic energy scale corrections $a^\prime_1$ and $a^\prime_2$, and no energy offset, implicitly expecting E=0 events to measure 0~ADC codes.
We find that this calibration curve accurately reconstructs peak positions from 0 to 2615~keV for our improved energy estimator, avoiding the offset at E=0 seen in our other approaches.
Approach~3 achieves a comparable energy resolution to the other approaches at 2039~keV, while correcting a shift of 0.15~keV (nearly one FWHM) in low energy peaks~\cite{Arnquist:2023ct}.
Furthermore, this approach rarely required human intervention to achieve stable fit results.

\section{Uncertainties}
\label{sec:uncertainties}
The energy calibration was applied to merge the individual detector spectra into one spectrum for the ~\nonubb\ analysis.  It was also used to calibrate the specialized cuts used in the the analysis to reject background events~\cite{Alvis:2019sil}\cite{Arnquist:2023}.  The calibration and particularly the peak fit hyper-parameters were used to determine the limits of the detection of ~\nonubb\ events in the region of interest.

The most significant limitations in detecting \nonubb\ were the uncertainty in the peak position, the shape of the peak, and the total efficiency for detecting an event at the expected energy.
The peak position mean was set by the calibration to the known values of the $^{228}$Th decay.   The statistical uncertainty ($\delta_{\mu}$) was determined from the calibration hyper-parameters.

This peak shape was used to determine the width of the energy window used to search for \nonubb\ events.  It was determined by the components of the peak fit, the Gaussian and the exponential tail. Because of the correlation of these contributions to the peak shape, a good approximation was to consider the worst possible deviation of the greatest contribution, which usually was the Gaussian contribution to the peak.   The relative uncertainty introduced by the peak shape was well approximated by the relative uncertainty of the FWHM, which was computed as a function of the gamma energy.

The total efficiency included several contributions.  The energy calibration contributed to the containment efficiency, which provided the fraction of \nonubb\ events that are detected in the energy window selected for the analysis. This fraction depends on the peak shape parameters: Gaussian peak width ($\sigma$), exponential tail fraction ($f$) and exponential tail slope ($\tau$).

These parameters were calculated for each dataset. The process to determine these variables  required a multi-peak fit using the combined calibration spectrum of the dataset. This fit used 18 peaks in the $^{228}$Th decay chain, from 215 keV to 2615 keV (figure \ref{fig:multipeakfwhm}) and fit $\mu$ for each peak without parameterization.  The values of $\sigma$ and $\tau$ were obtained from the hyper-parameters while $f$ was obtained only from the 2615 keV peak.  This peak was chosen because it had the highest statistics among those close to the \nonubb\ position.  The FWHM and its uncertainty were also determined from this fit.  Changes in the gain match parameter, Eq.~\ref{eq:gainmatch}, had a small but significant effect in the energy window. To include this effect, we calculated the change of the uncalibrated peak position between adjacent calibrations as

\begin{linenomath*}
\begin{align}
   \Delta_{i,j} \left(keV\right)= \left(E^{2614.5keV}_{i,j}-E^{2614.5keV}_{i,j+1}\right)a_{i,j}
   \label{eq:eshift}
\end{align}
\end{linenomath*} 
where
\begin{itemize}
\item $E^{2614.5keV}_{i,j}$ = energy position, in ADC units, of the 2614.5 keV peak in the detector \textit{i} and the calibration \textit{j}.
\end{itemize}

If this parameter was more than 2 keV for an energy calibration, the data for the channel between calibration $j$ and $j+1$  was not analyzed.  This occurred for $2.8\%$ of the data.  Using the analyzed data, this contribution to the FWHM was defined as $FWHM_{gain} = 2.355\times\sigma_{\Delta_{i,j}}$/2, using the standard deviation of the peak position fluctuations.  

The uncertainty in the peak position, $\delta_{\mu}$,  in addition to its statistical uncertainty obtained from the fit, had three non-statistical contributions: the global effect of the calibration non-linearity, the gain fluctuations, and the ADC non-linearity~\cite{Abgrall:2020jto}.  
The statistical uncertainty was determined from the energy scale in the fit with the 18 peaks.  A linear function, $\mu$ = $\mu_{0}$ + $\mu_{1}E$, was used to fit the peak positions to their nominal values.  If the energy calibration was successful, $\mu_{0}$ was comparable to 0 and $\mu_{1}$ was comparable to 1 within errors. The uncertainties of those magnitudes were used then to determine the statistical error as

\begin{equation}
\delta_{stat}\left(E\right) = \sqrt{\delta^{2}_{\mu_{0}}+\delta^{2}_{\mu_{1}}E^{2}+2E \times\! cov(\mu_{0},\mu_{1})}
\end{equation} 
where
\begin{itemize}
\item $\delta_{\mu_{0}}$ = uncertainty of $\mu_{0}$ obtained from the fit.
\item $\delta_{\mu_{1}}$ = uncertainty of $\mu_{1}$ obtained from the fit.
\item $cov(\mu_{0},\mu_{1})$ = covariance of  $\mu_{0}$ and $\mu_{1}$.
\end{itemize}

However, this uncertainty was not used alone to calculate $\delta_{\mu}$ because the global non-linearity uncertainty can not be determined by itself.  Instead, we computed an uncertainty term that combined both statistical and global non-linearity (gNL). To estimate the effect of all the non-linearities in the uncertainty, we presumed that the non-statistical residuals in the peak locations were due to the non-linearity.  We rescaled the statistical variance, determined from the 18 peak fit parameters, by the ratio of $\chi^2$ to the number of degrees of freedom:

\begin{equation}
\delta_{stat+gNL}(E) \approx \delta_{stat}(E)\sqrt{\frac{1}{N_{p}-2}\sum_{k=1}^{N_{p}}\frac{\Delta_{k}^{2}}{\delta^{2}_{stat,k}}}
\end{equation}
where
\begin{itemize}
\item $N_{p}$ = number of peaks.
\item $\Delta_{k}$ = difference between the $\mu$ value and the energy nominal value of the peak \textit{k}.
\item $\delta_{stat,k}$ = $\delta_{stat}(E)$ at the nominal energy of the peak \textit{k}.
\end{itemize}

The last uncertainty contribution was due to fluctuations in the gain, Eq.~\ref{eq:gainmatch}.  It was calculated as
\begin {equation}
\delta_{drift} =  \sqrt{\frac{\sigma^{2}_{drift}}{N}+\left(\frac{1}{2}\overline{\Delta}_{i,j}\right)^{2}}
\end{equation}
where $N$ is the number of periods  accepted for the analysis.  This equation takes the standard deviation of 
$\Delta_{i,j}$ as the average uncertainty introduced by the change in the gain. 
However, the fact that the change was not the same between any pair of adjacent calibrations necessitates adding a term to take into account the different magnitudes of the gain change.   If the fluctuations have a consistent drift with small variance, a better estimate of the uncertainty was half the average fluctuation.  We used the sum to conservatively estimate the effect.

Calculating the energy parameter for each event and estimating the systematics for the statistical analysis completed the energy calibration procedure.

\begin{figure}[hb]
\centering
\includegraphics[width=\columnwidth]{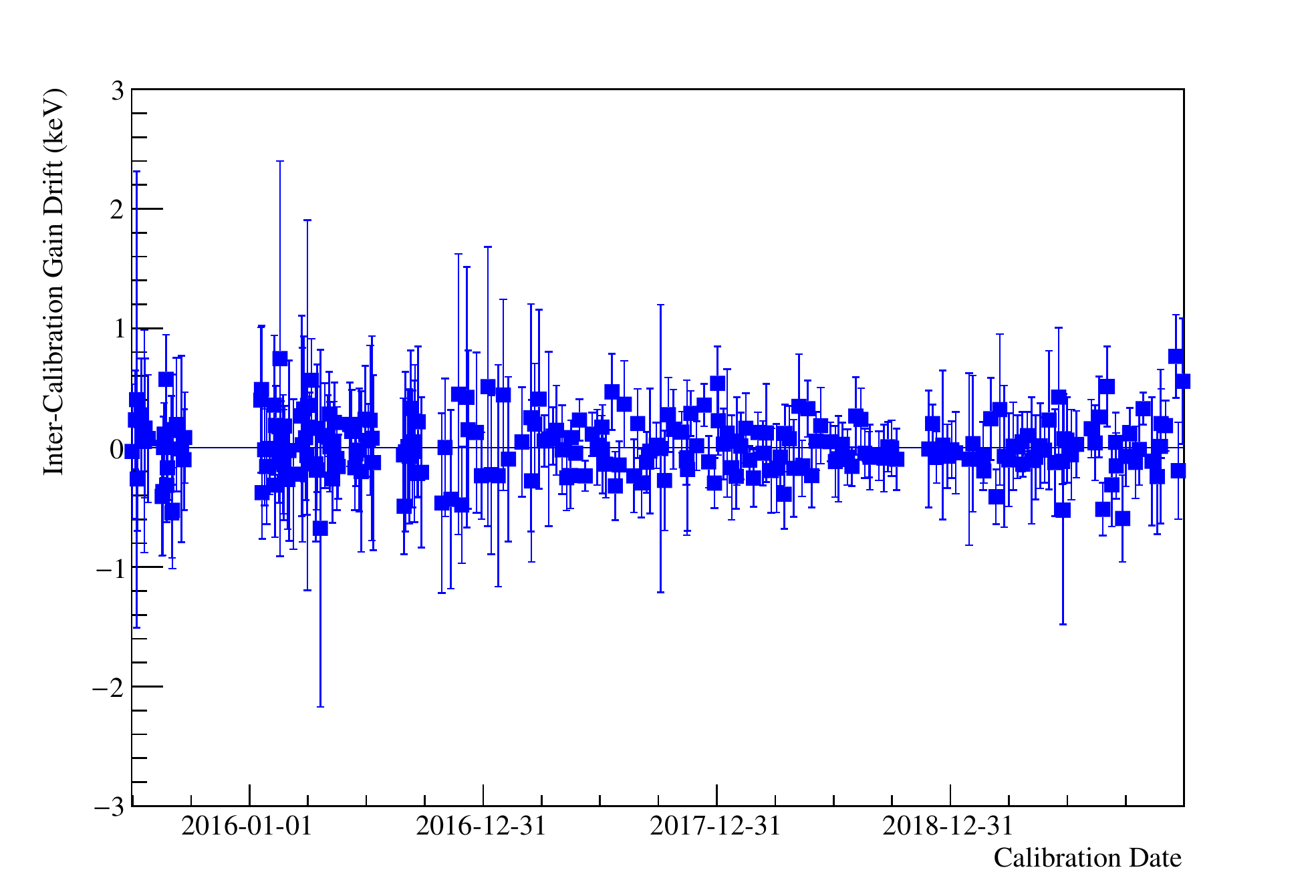}
\caption{The energy deviation of the 2615-keV peak between two contiguous calibration periods of all operating high-gain detectors in the \Dem~\nonubb~search reported in Ref~\cite{Aalseth:2017,Alvis:2019sil}. The standard deviation of the gain drift in each data set is about 0.5 keV. The big gap before 2016-01-01 is because of downtime for final configuration of the experiment. The uncertainty became smaller after 2016-12-31 because of the improvement of grounding and the completion of construction.}
\label{fig:delta_bar}
\end{figure}

\section{Conclusion}
\label{sec:conclusion}
The \Dem~has achieved the lowest energy resolution of $2.52\pm0.08$ keV FWHM at 2039 keV~\cite{Arnquist:2023}, leading among \nonubb~experiments. This has been achieved with PPC detector technology, low-noise readout electroncs, the ADC NL correction, the detector charge trapping correction and the calibration procedure described here. In this paper, we have described the offline procedure of energy calibration for the \Dem~experiment. The peaks in the $^{228}$Th energy spectrum of each detector were well modeled by a multi-peak fitting algorithm. The multi-peak fitter used an HMC method to estimate its initial parameters for a gradient minimization algorithm, including the peak shape and the background shape parameters. 

\section{Acknowledgements}
This material is based upon work supported by the U.S.~Department of Energy, Office of Science, Office of Nuclear Physics under contract / award numbers DE-AC02-05CH11231, DE-AC05-00OR22725, DE-AC05-76RL0130, DE-FG02-97ER41020, DE-FG02-97ER41033, DE-FG02-97ER41041, DE-SC0012612, DE-SC0014445, DE-SC0018060, DE-SC0022339, and LANLEM77/LANLEM78. We acknowledge support from the Particle Astrophysics Program and Nuclear Physics Program of the National Science Foundation through grant numbers MRI-0923142, PHY-1003399, PHY-1102292, PHY-1206314, PHY-1614611, PHY-1812409, PHY-1812356, PHY-2111140, and PHY-2209530. We gratefully acknowledge the support of the Laboratory Directed Research \& Development (LDRD) program at Lawrence Berkeley National Laboratory for this work. We gratefully acknowledge the support of the U.S.~Department of Energy through the Los Alamos National Laboratory LDRD Program and through the Pacific Northwest National Laboratory LDRD Program for this work.  We gratefully acknowledge the support of the South Dakota Board of Regents Competitive Research Grant. 

We acknowledge the support of the Natural Sciences and Engineering Research Council of Canada, funding reference number SAPIN-2017-00023, and from the Canada Foundation for Innovation John R.~Evans Leaders Fund.  We acknowledge support from the 2020/2021 L'Or\'eal-UNESCO for Women in Science Programme.  This research used resources provided by the Oak Ridge Leadership Computing Facility at Oak Ridge National Laboratory and by the National Energy Research Scientific Computing Center, a U.S.~Department of Energy Office of Science User Facility. We thank our hosts and colleagues at the Sanford Underground Research Facility for their support.

\bibliographystyle{unsrt}
\bibliography{reference}
\end{document}